\documentstyle[twocolumn,seceq]{jpsj}

\def\runtitle{ESR study on the excited state energy spectrum of SrCu$_2$(BO$_3$)$_2$-a central role of multiple-triplet bound states}

\def\runauthor{Hiroyuki {\sc Nojiri}, Hiroshi {\sc Kageyama}, Yutaka {\sc Ueda}, Mitsuhiro {\sc Motokawa}}

\title
{ESR study on the excited state energy spectrum of SrCu$_2$(BO$_3$)$_2$\\-a central role of multiple-triplet bound states}

\author
{Hiroyuki {\sc Nojiri}, Hiroshi {\sc Kageyama}$^1$, \\Yutaka {\sc Ueda}$^1$ and Mitsuhiro {\sc
Motokawa$^2$}
}

\inst
{Department of Physics, Okayama University, Tsushimanaka 3-1-1, OKayama 700-8530

$^1$Institute for Solid State Physics, University of Tokyo, Kashinoha 5-1-5, Kashiwa 277-8581

$^2$Institute for Materials Research, Tohoku University, Katahira 2-1-1, Sendai 980-8577
}

\recdate
{
\
}

\abst
{
The excited state energy spectrum of a two-dimensional dimer system SrCu$_2$(BO$_3$)$_2$ has been investigated by high field ESR.
Various types of multiple-triplet bound states-excited singlet, triplet and quintet states are identified, besides two non-degenerated 
one-triplet excitations.
Extremely strong intensity is found for these higher order bound states, which is an important consequence of the strong suppression of the one-triplet
process due to the orthogonality of dimers.
A new bound singlet state is found slightly below the energy gap of the
one-triplet state, indicating the proximity to a quantum critical point.
A band of two-triplet bound state splits into at least eight sharp and discrete levels, which indicates the contribution of distant neighbor
interactions.
A bound quintet state is observed, which is the direct evidence of two-triplet object made up of four $S=1/2$ spins.
The energy of the lowest quintet becomes lower than that of the one-triplet excitation before the closing of the lowest spin gap.
It indicates that the two-triplet bound state touches to the ground singlet state before the one-triplet state does.
The present investigation exhibits an essential role of various multiple-triplet bound states in the magnetic excitation spectrum of
SrCu$_2$(BO$_3$)$_2$.}

\kword
{
SrCu$_2$(BO$_3$)$_2$, spin gap, ESR, magnetic excitation, high
magnetic field }

\begin{document}
\sloppy
\maketitle

\section {Introduction}
Coupled chain and dimer systems have been attracted much attention in this decade.
One of the interests is in a complex and fruitful crossover from a
well-known purely one-dimensional antiferromagnet to a two-dimensional(2D) antiferromagnet, i.e., the  basis of a high-temperature
superconductor.\cite{rice}
Experimentally, a 2D spin gap system is quite rare, in contrast with a rapid growth of new quasi-one dimensional compounds.
One of fundamental difficulties in studying the 2D system experimentally is that it tends to exhibits three dimensional(3D) long range
order. However, there is an effective way to prevent the 3D-order:introduction of frustration into a system. 
One of the most successful examples is SrCu$_2$(BO$_3$)$_2$:a 2D dimer system with considerably strong frustration.\cite{kage}

In this compound, a layer of $S=1/2$ Cu$^{2+}$ ions is sandwiched by layers of Sr ions in a tetragonal unit cell.\cite{smith}
A dimer unit is composed of a  neighboring pair of edge-shared planar rectangular CuO$_4$ and these dimers connect
orthogonally by way of a triangular planar BO$_3$, providing a unique 2D-network as shown in Fig. 1.
This 2D lattice is topologically equivalent to a 2D square lattice with additional alternating diagonal interactions, for which the direct product of the
singlet pairs is the exact ground state of the system, as proven by Shastry and Sutherland.\cite{ss}

Although the ground state is known rigorously as a dimer solid, the excited state is expected to be very rich.
A low-lying excited triplet, which is dominant for spin gap compounds in general, is considerably suppressed due to the extreme localization.\cite{mu}
Higher order multiple-triplet states are instead stabilized by the kinetic energy called "correlated hopping". 
Those interesting features of the excited state have been studied experimentally and
theoretically, although a complete understanding has not been obtained so far.\cite{weihong,nd,pl,nojiriscbo,kn,mtmag,room,mt,fuk,nojiriesr,cepas}

Another important issue is that this compound is located very close to a quantum critical point.
The ratio $J_2$/$J_1$ is evaluated to be 0.63-0.68,  where $J_1$ and $J_2$ are, respectively, the intradimer and interdimer interactions as shown in Fig. 1
(a).\cite{mu3D} The value is just below the critical value of
$J_2$/$J_1$=0.70 between the dimer state and the N${\acute e}$el-ordered state. More recently, it has been proposed that a novel ground state, a resonating
plaquette singlet state, exists in between these two states.\cite{kk} Although the present compound is in the dimer phase, a precursor effect of other
types of ground states may appear, especially in excited states, for the proximity to the quantum critical point.

An interesting feature also appears in the magnetization curve with quantized plateaux at one-third, one-quarter and
one-eighth of the Cu saturated moment.\cite{magissp,magosaka}
These plateaux are caused by the extreme localization of the triplet excitation.
Namely, the excited triplets cannot propagate freely in the 2D lattice owing to the orthogonal dimer arrangement and consequently organize a regular lattice
composed of spatially separated triplets when the density of the triplets is commensurate with the underlying lattice. 

As discussed above, it is very interesting to study the excited state spectrum of the present system, especially as a function of a magnetic field.
Among many different methods, ESR has established its unique status as a probe of magnetic excitations, particularly in studying the high field
property.\cite{cugeo,nav,nh4,cbz}
In fact, the one-triplet as well as the multiple-triplet excitations were reported in our previous paper.\cite{nojiriscbo}
However, the richness of the spectrum and the limited frequency range prevented us to complete the experimental determination of a full spectrum.
In the present paper, a nearly complete excited state energy spectrum of the SrCu$_2$(BO$_3$)$_2$ up to the two-triplet continuum and up to
40 T is obtained.

The format of the paper is as follows.
After briefly mentioning the experimental procedure, we show details of experimental results.
The energy diagram, selection rule and the symmetry of one-triplet and multiple-triplet excitations are examined.
The anomalous behavior of the magnetic excitation and its relation to the magnetization around the critical field is discussed in terms of the
localization of the bound states.
\begin{figure}
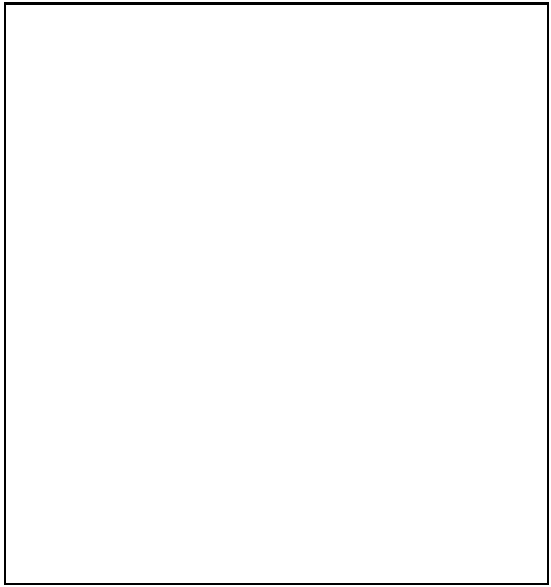

\figureheight{7.5cm}
\caption{2D-dimer plane viewed along the $c$-axis with respect to z=0, or 0.5. Open, shadowed and closed circles, denotes Cu, O and B ions,
respectively. The size of a circle indicates the shift along the $c$-axis(bigger/smaller, corresponding upward/downward). An arrow on Cu shows a tilting of
the principle axis of
$g$-tensor. A and B distinguish two types of dimers(with dashed ellipsoids) which are connected orthogonally each other. Thin square shows a unit cell. 
$J_1$(thick dashed line) and  $J_2$(thin dashed line) are the nearest neighbor and the next-nearest-neighbor interaction in the plane, respectively. The
inset is the [1-10]-axis view of the A-dimer. Ellipsoids schematically represent the anisotropy of $g$-tensor.}
\label{fig:1}
\end{figure}
\section{Experimental procedure}

 Submillimeter wave ESR measurements have been performed up to 1.3 THz and in pulsed magnetic fields up to 40 T.
An optical pumped far-infrared laser, backward-travelling wave tubes and Gunn oscillators have been employed as
radiation sources.
We have employed a simple transmission method with Faraday configuration where the propagation vector of the incident radiation is aligned
parallel to the external magnetic field.
The polarization of light is random in the plane normal to the propagation vector.
An InSb is used as the detector. 
The detail of the ESR system is given in the references.\cite{nojiriesr,nojirioka}

High-purity bulk single crystals of SrCu$_2$(BO$_3$)$_2$ were grown by the travelling
solvent floating zone (TSFZ) method from a polycrystalline SrCu$_2$(BO$_3$)$_2$ and a solvent LiBO$_2$ using FZ-T10000N 10KW high
pressure type (Crystal System Inc.) with four halogen lamps as heat sources.
Oxygen gas (P$_{O2}$=1 atom, 99.999 $\%$) was being flowed during the growth process.\cite{cryst}
By means of Laue X-ray back-reflection, the grown materials were checked and the crystal axes were determined. A piece of the single crystal with the
dimensions of about 4 mm${\times }$4 mm${\times }$1 mm was used for the ESR experiments.

\section{One-triplet excitation}
In the present work, we mainly focus on ESR spectra at 1.6 K and below the critical field $H_c$ at which the gap closes.
Since this temperature is much lower than the spin gap, all the signals are assigned to be the excitations from the ground state.
In this section, we show important features of an one-triplet excitation:a creation of single triplet in a dimer, and discuss the ESR
selection rule considering Dzyaloshinsky-Moriya(DM) interactions and a staggered field. 
To avoid confusion, we attach subscripts:GS(ground state) and BS(bound state or multiple-triplet excitation) to
the terms, singlets and triplets. For examples, singlet$_{GS}$, singlet$_{BS}$ and triplet$_{BS}$ represents the singlet ground state, the singlet
excited bound state and the triplet bound state, respectively.

\subsection{Experimental features}
\begin{figure}
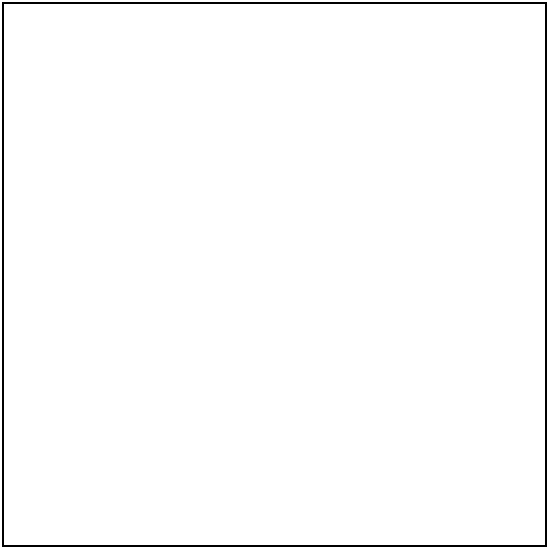

\figureheight{7cm}
\caption{Temperature dependence of ESR spectra at 428.6 GHz (a) for {\mib {B}}${\parallel}c$ and (b) for {\mib {B}}${\parallel}a$.
The sharp peak at 15.3 T is the signal of DPPH used as a field maker. The inset shows a schematic energy diagram of
triplet with a spin gap.}
\label{fig:2}
\end{figure}
Figure 2 (a) shows a typical temperature dependence of ESR spectrum.
The intensity of a broad peak at 13.3 T decreases rapidly below 10 K.
It indicates that the peak is caused by the transitions of the thermally excited triplets as shown by dotted arrows in the
inset. At 1.6 K, two weak signals marked by arrows are observed.
Since the intensity of the weak signals increases with decreasing temperature, these are assigned as the transitions between the ground singlet and the
excited states, which is indicated by the solid arrows in the inset. While the similar temperature dependence is found for {\mib
{B}}${\parallel}a$ in Fig. 2 (b), differences are found in the splitting and the intensity of the two peaks. As is well known, an ESR transition between the
ground singlet and the excited triplet states due to the mechanism of magnetic dipolar transition is usually forbidden. \cite{sakai}
A non-secular term such as DM-interaction, as proposed recently by C$\acute e$pas {\it et al.}, may be the leading origin of the break down of the
selection rule.\cite{cepas} This point will be discussed later.
\begin{figure}
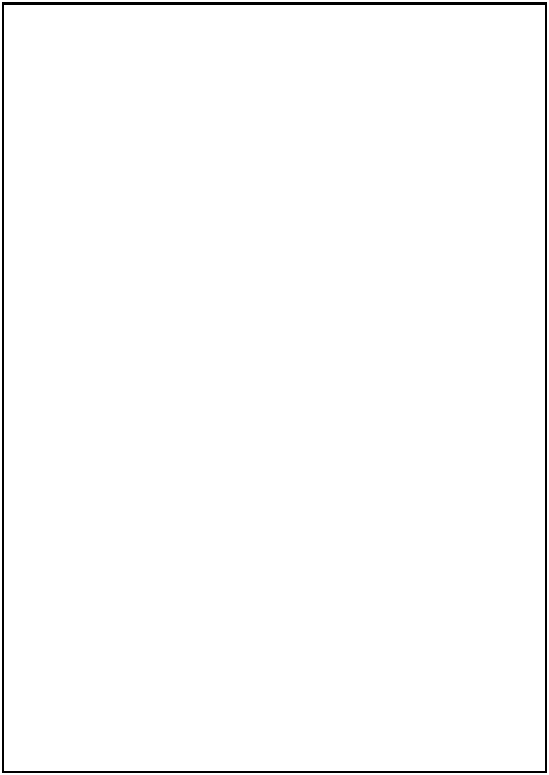

\figureheight{10cm}
\caption{Summarized magnetic excitation spectra spanning 0-40 T and 0-1100 GHz at 1.6 K, (a) for {\mib {B}}${\parallel}c$, (b) for
{\mib {B}}${\parallel}a$. To reduce the data volume, the data points are averaged with the interval of 0.1 T.}
\label{fig:3}
\end{figure}

In Figs. 3 (a) and 3 (b), ESR spectra measured at different frequencies are collected together for {\mib {B}}${\parallel}c$ and
{\mib {B}}${\parallel}a$, respectively. In these figures, each ESR spectrum is plotted in such way that its background coincides with the measured
frequency, so that the resultant plot can be regarded as a kind of the frequency-field diagram which furthermore show intensity variation of
excitations. A detailed frequency field diagram is obtained by tracing different types of peaks in the plot. Frequency-field
diagrams thus obtained are depicted in Figs. 4 (a) and 4 (b) for {\mib {B}}${\parallel}c$ and for {\mib {B}}${\parallel}a$, respectively. A contour map of
the intensity in the frequency-field plane as shown in Fig. 5. Those figures are useful to see the overview of the excited state energy spectra in
field-frequency plane. A complicated structure of the spectrum is noticed, that cannot be expressed well by a simple frequency-field plot. In the following,
we show important features of the one-triplet excitation using the set of these figures.

In Figs. 4 (a) and 4 (b), it is noticed that two low-lying triplet excitations:O$_1$ and O$_2$ exist.
The values of the zero field gaps are close to the spin gap estimated from magnetic susceptibility and high field magnetization{\cite{kage,magissp} and
thus these modes are assigned as one-triplet excitation.
More precisely, the spin gap $E_g$ is determined to be 722$\pm$2 GHz, which is evaluated as the average of the zero field intercepts of O$_1$ and O$_2$. 
The splitting between O$_1$ and O$_2$ is largest for {\mib {B}}${\parallel}c$ and the zero field energy gaps of O$_1$ and O$_2$ are 764$\pm$2 GHz and
679$\pm$2 GHz, respectively.
The existence of two modes is generally expected when a unit cell contains two magnetic ions.
The two modes, originally degenerated, may be split by an anisotropic inter-dimer interaction and/or other perturbations.
In the present compound, the splitting of the two-modes is interpreted fairly well by including a
inter-dimer DM-interaction.\cite{cepas}

When a magnetic field is rotated from the {\mib {c}}-axis to the {\mib {a}}-axis, the splitting shows a monotonic decrease
 and takes a non-zero minimum for {\mib {B}}${\parallel}a$ as we reported in the previous work(see Fig. 4 of Ref.[15]).
The angular dependence of the upper-going branches of O$_1$ and O$_2$ is more clearly shown in Fig. 6.
The two modes are almost parallel to each other for {\mib {B}}${\parallel}c$, while they become very close each other for {\mib {B}}${\parallel}a$ accompanied by the bending around
zero field. These behaviors indicate that the principle axis of the interaction causing the splitting points to the {\mib {c}}-axis. 
\begin{figure}
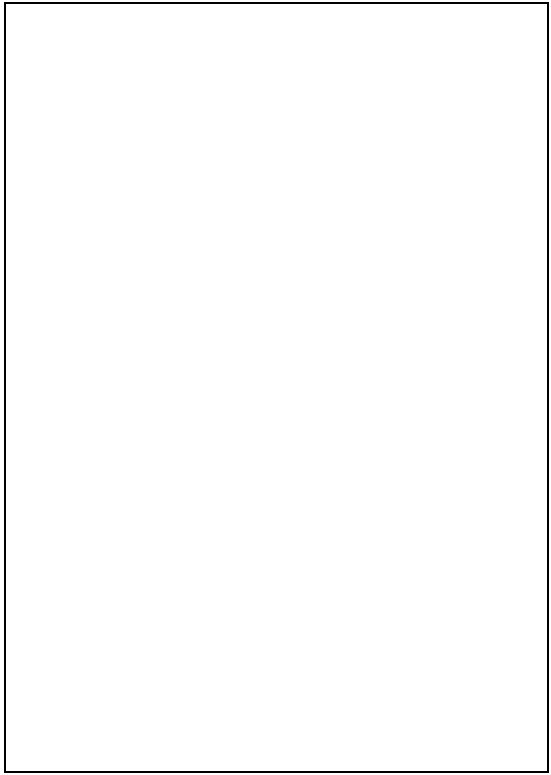

\figureheight{10cm}
\caption{Frequency field diagram at 1.6 K (a)for {\mib {B}}${\parallel}c$ and (b) for {\mib
{B}}${\parallel}a$. O, T, Q, B, M are one-triplet, triplet$_{BS}$, quintet$_{BS}$, new signal and paramagnetic like signal, respectively. For
details, see text. The subscript W denotes weak signals which are not well classified. Solid line shows the paramagnetic resonance line(${\nu=g\mu_BH}$,
$\nu$:frequency,
$g$:$g$-value). Dashed lines are eye guides. Magnetization curves taken from Ref.[20] are plotted together.}
\label{fig:4} 
\end{figure}
\begin{figure}
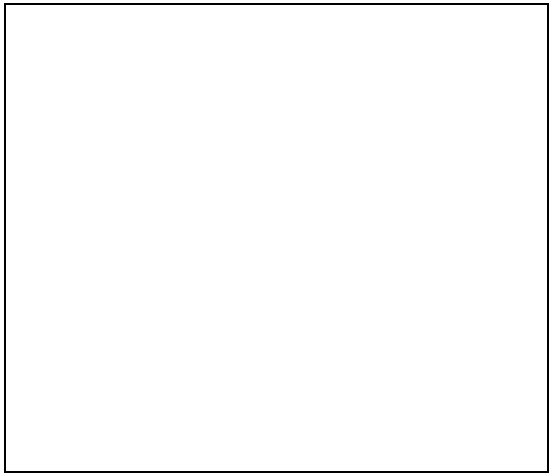

\figureheight{6cm}
\caption{Contour plot of ESR absorption intensity for {\mib {B}}${\parallel}a$ at 1.6 K. The color scale is arbitrary.}
\label{fig:5}
\end{figure}
\begin{figure}
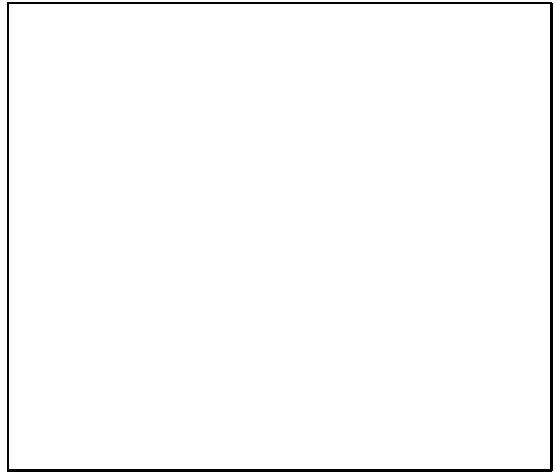

\figureheight{6cm}
\caption{Angular dependence of O$_1$ and O$_2$ in the $ac$-plane. $\theta$ is the angle between the $a$-axis and an external field. Lines are
eye-guides. Anisotropy of 
$g$-value is plotted in the inset, which is measured at 25 K with 135 GHz. The negative angle corresponds to the rotation of an external field from the
$a$-axis to the $b$-axis.}
\label{fig:6}
\end{figure}

It should be stressed here that a small but intrinsic splitting exists between O$_1$ and O$_2$ for {\mib {B}}${\parallel}a$.
It should be noted that the splitting is not caused by the sample misorientation or anisotropy of $g$-value. 
This can be partly confirmed if one measures the two modes for {\mib {B}}$\parallel$[110]. In this case, as
shown in Fig. 1, the magnetic field is parallel to the A-dimers and is perpendicular to the B-dimers. A larger splitting would be expected, if it was caused
by the difference of
$g$-value between the parallel and the perpendicular directions of a dimer. However, it is found that the splitting is nearly identical with that for {\mib
{B}}${\parallel}a$(not shown here). The result indicates that the splitting observed for {\mib {B}}${\parallel}a$ is not caused by the $g$-value
anisotropy. 

The essential difference between O$_1$ and O$_2$ in {\mib {B}}${\parallel}a$ is found in the following mixing behaviors.
Firstly, it is found that the magnitude of the mixing with singlet$_{BS}$ states is different between O$_1$ and O$_2$. In
Fig. 4 (b), O$_1$ shows a break around 860 GHz, while O$_2$ shows another break around 650 GHz. These breaks may be caused by an anti-level-crossing
with an excited singlet state(this point will be discussed later). Secondly, it is noticed that O$_1$ and O$_2$ round very differently around $H_c$
where a mixing with the ground state occurs. The mixing is strongly related to the symmetry of corresponding modes, and thus these
findings indicate that the symmetry of O$_1$ is different from that of O$_2$. 

Next we discuss the intensity of the one-triplet excitation.
For {\mib {B}}${\parallel}c$, a clear difference is found in the field dependence of intensity between O$_1$ and O$_2$ as shown in Fig. 3 (a).
The intensity is nearly field independent for O$_1$, while it is sensitive to the magnetic field intensity for O$_2$.
Namely, the intensity of the down-going branch of O$_2$ increases with increasing field and that of the up-going branch shows an opposite field dependence.
In addition to this, a rapid decrease of intensity is observed for both branches around zero field. 
The behavior suggests that the origin of the break down of the selection rule is different between O$_1$ and O$_2$ in {\mib {B}}${\parallel}c$.
For {\mib {B}}${\parallel}a$, the behavior of O$_1$ and O$_2$ is much different as follows.
First, the intensity is nearly field independent for both O$_1$ and O$_2$.
Secondly, the intensity is stronger than that in {\mib {B}}${\parallel}c$.
The second feature suggests that a non-secular term causing these transitions should be larger for {\mib {B}}${\parallel}a$.

We compare the present result with that of the FIR spectroscopy\cite{room}.
It should be reminded that the definitions of absorption intensity are not exactly same each other.
In our case, the intensity is normalized by the intensity of DPPH used as maker or by the intensity of paramagnetic signal.
In the FIR experiment, the magnetic intensity is obtained by taking a difference between a high temperature spectrum and a low temperature one.  
In spite of this difference, two results are roughly consistent each other.
In Fig. 4 of Ref. [12], a field is applied along the $a$-axis.
The intensity decreases with increasing field for the $c$-polarization, while it increases for the ${ab}$-polarization. 
In case of ESR, un-polarized radiation is used and thus the intensity should be averaged over the two polarizations.
If we average the data of different polarizations obtained by FIR spectroscopy, it can happen that the intensity depends on magnetic field only weakly
for the cancellation, which agrees with the ESR result for {\mib {B}}${\parallel}a$.

\subsection{Selection rule and DM-interaction}

In the following, we discuss possible interactions allowing the finite intensity of the one-triplet excitations, which is originally forbidden.
We consider
an intra-dimer DM-interaction: {\mib {D}$_d$}, the $c$-axis and the ${ab}$-plane components of inter-dimer DM-interaction: ({\mib {D}$_c$ and {\mib
{D}$_{ab}$, respectively) and a staggered field, all of which are allowed to exist by its crystal structure. We do not consider a novel
mechanism:dynamical DM-interaction proposed recently.\cite{cepasFJ,cepasESR} Now the point to be clarified is what aspect of the
experimental result can be interpreted with those four terms. We would refuse to deny or confirm the existence of the dynamical DM-interaction
because the judgement cannot be made with present experimental results. 

We employ the selection rule established by Sakai, C$\acute e$pas and Ziman for DM-interaction.\cite{sakai}
Since the present experiment is made in Faraday configuration, one simple rule is applicable as follows.

DM-rule:{\it Field independent intensity is observed only when the DM-interaction has a component along the external magnetic field and the intensity is
quadratic to the component.}

Among three possible DM-interactions, {\mib {D}$_c$ exists irrespective of the small buckling of the edge-shared CuO$_4$ planes as shown in Fig.
1. {\mib {D}$_d$} and {\mib {D}$_{ab}$ exist only when this buckling is taken into account.
At the same time, a staggered field becomes possible for the alternation of a principle-axis of $g$-tensor as shown in the inset of Fig. 1.
Considering the large anisotropy of $g$-value in the ${ac}$-plane(${g}_a$=2.05 and ${g}_c$=2.28) and the small one in the ${ab}$-plane($\Delta g \le$ 0.01,
see Fig. 6 inset), a sizable staggered field is expected only for {\mib {B}}${\parallel}c$. Those three terms caused by the buckling may be small, because of
the small tilting angle. An experimental estimate of {\mib {D}$_{ab}$ is made by Kakurai {\it et al.}\cite{KakFJ} and it is found that the magnitude of {\mib
{D}$_{ab}$ is about 30 $\%$ of that of {\mib {D}$_c$. This estimation, however, looks overestimated since the buckling angle is quite small. The important
point is that the practical effect is observed with such small buckling and thus {\mib {D}$_c$ cannot be neglected. 
It should be also noted that {\mib {D}$_d$} is much effective than {\mib {D}$_{ab}$ and {\mib {D}$_c$.
It is simply because {\mib {D}$_d$} operates directly on
the two spins in a dimer. From the crystal symmetry it is evident that {\mib {D}$_d$} lays on the
${ab}$-plane and is perpendicular to the direction of a dimer bond. 
Since two kinds of dimers are orthogonally arranged in the ${ab}$-plane, the sum of the squares of a field-parallel-component of
{\mib {D}$_d$} over the A-dimers and the B-dimers is constant for any field direction in the plane.

Let us examine, if the DM-rule is compatible with the experimental results.
Considering the DM-rule, {\mib {D}$_c$ gives rise to constant ESR intensity for {\mib {B}}${\parallel}c$.
This expectation is consistent with the behavior of O$_1$ in which the intensity is field-independent.
A precise estimation of {\mib {D}$_c$=0.18 meV is made from the dispersion measured by neutron scattering.\cite{cepas}
This value is a standard when we estimate the magnitude of DM-interaction from the ESR intensities.  

Contrary to O$_1$, the field dependence of the intensity of O$_2$ is not compatible with the DM-rule and thus DM-interaction is not the origin
of the transition O$_2$.
As shown in the previous section, the intensity of O$_2$ shows the characteristic field dependence.
It is noticed that the behavior is similar to the case of a
staggered field with Voight configuration\cite{shiba}. 
In fact, a staggered field is expected for this field orientation as mentioned above.
However, it cannot be applicable to the present experiment because the Faraday configuration is used.
It is clear that neither DM interaction nor a staggered field can cause the transition O$_2$ in the lowest order from the selection rules.
A possibility is that a combined higher order term may cause a finite ESR intensity.
The fact that the intensity of O$_2$ is much weaker than that of O$_1$ is consistent with the contribution of the higher order term.
A similar situation is found in CuGeO$_3$, in which both optical and acoustic branches are observed in ESR.
This behavior cannot be interpreted by the lowest order term of DM-interaction.\cite{sakai}
As discussed above, it is rather plausible that a combined higher order term of DM interaction and a staggered field gives the transition O$_2$.
A theoretical calculation of the intensity is necessary to clarify the present proposal.

Next we discuss the case of {\mib {B}}${\parallel}a$.
Since both {\mib {D}$_d$ and {\mib {D}$_{ab}$ have the field parallel components, the field independent intensity observed in the experiment is
compatible with the DM-rule.
The fact the intensity is larger in {\mib {B}}${\parallel}a$ than that in {\mib {B}}${\parallel}c$ indicates that the field parallel DM-component is
larger in {\mib {B}}${\parallel}a$. As mentioned above, the square of the magnitude of {\mib {D}$_{ab}$ is one-order smaller than that of {\mib
{D}$_{c}$ and thus {\mib {D}$_{ab}$ causes the minor contribution to the intensity. Hence, {\mib {D}$_d$ should be dominant in {\mib {B}}${\parallel}a$.
Another piece of the evidence of the major contribution of {\mib {D}$_d$} is that the intensity of O$_1$ and O$_2$ is almost constant
when we rotate an external magnetic field within the ${ab}$-plane(not shown here).
It is consistent with the fact that the sum of the squares of the field parallel components of {\mib {D}$_d$} in A and B-dimers is constant under this
field rotation. 
Moreover, {\mib {D}$_d$} operates directly on each dimer, while {\mib {D}$_{ab}$ does not. 
All these facts indicate that the {\mib {D}$_d$} is the main cause of the selection rule breaking in {\mib {B}}${\parallel}a$. 

Finally, we would like to make a comment on an exchange anisotropy. In our previous work, we did not consider the inter-dimer DM-interaction and
mentioned that exchange anisotropy may cause the splitting between O$_1$ and O$_2$ and the forbidden transitions.
Recent proposal by C$\acute e$pas {\it et al.} that DM-interaction is the leading term is more plausible than the exchange anisotropy. Now we consider that
a simple exchange anisotropy term does not cause the mixing between the singlet ground and the excited triplet states. However, a consideration of the
problem from more general view point may be interesting.\cite{Kap,SWA,smith} High resolution of ESR can resolve even a small change of triplet
excitation caused by these terms.

\section {Multiple-triplet bound states}
As shown in Figs. 3 (a) and 3 (b), many high energy excitations, which have been interpreted as multiple-triplet bound states,\cite{nojiriesr} account for
a considerable weight in the excited state.
In the present system, a sizable inter-dimer interaction exists.
This interaction couples two(or three) dimers and causes a bound sate with total spin {\it S}=2}(or {\it S}=3).
A {\it S}=2} bound state, for example, consists of three sub-states;singlet, triplet and quintet.
Usually such bound state appears as a continuum.
In an orthogonal-dimer system, however, discrete states are expected for the localization of triplets. 
In this section, we describe the behaviors of those bound state excitations in detail.

\subsection {Singlet bound state}
A singlet$_{BS}$ is an excited singlet state  consisting of two interacting triplets on nearby dimers and the energy is given by subtracting the binding
energy from the twice of $E_g$.
For a frustrated spin gap system, it is expected theoretically that excited singlets can dominate the states below $E_g$.
In highly frustrated system, an excited singlet lies in the low energy range close to the ground state.
In another word, the magnitude of the binding energy is a yardstick for the degree of frustration.
In the present case, such a low energy excited singlet indicates the close distance from a quantum critical point.

In the present work, an anti-level-crossing between a singlet$_{BS}$ and a one-triplet excitation is found
as shown in Figs. 7 (a) and 7 (b).
In Fig. 7 (a), as the frequency increases toward 646 GHz, O$_2$ deviates from O$_1$ and shifts to the low-field side with a considerable broadening. 
Above this frequency, O$_2$ approaches to O$_1$ from the high-field side.
For O$_1$, a conventional successive Zeeman shift is found.
The bending of O$_1$ around zero field is caused by the anisotropic inter-dimer DM-interaction as discussed in the previous section.
The break of O$_2$ shown in Figs. 7 (a) and 3 (b) is a typical behavior of an anti-level-crossing.
Although the singlet$_{BS}$ cannot be observed directly by ESR, the anti-level-crossing exhibits clearly the existence of the
singlet state TS$_1$ at 646 GHz. 
 
A similar anti-level-crossing is found at 860 GHz between the singlet TS$_2$ and O$_1$ as shown in Fig. 7 (b).
When the frequency is very close to TS$_2$, O$_1$ disappears and only a single peak of O$_2$ is found. 
The magnitude of the anti-level-crossing at TS$_2$ is larger than that at TS$_1$.
It indicates that the coupling between the singlet$_{BS}$ and the one-triplet state is larger in TS$_2$. 
The present result also exhibits that the symmetrys of the one-triplet excitations O$_1$ and O$_2$ are different each other as mentioned in the previous
section. In Fig. 7 (c), a frequency-field diagram is plotted, where a magnetic field is in the $ab$-plane but not parallel to the $a$- nor
the $c$-axes. An anti-level-crossing is found in both O$_1$ and O$_2$ at both TS$_1$ and TS$_2$(four breaks in total).
This behavior shows that two-modes O$_1$ and O$_2$ are mixed each other when an external field is tilted from the principle-axes($a$ or $c$) of the
crystal. It implies that the anti-level-crossing is closely related to the anisotropic part of exchange coupling such as DM-interaction.  
\begin{figure}
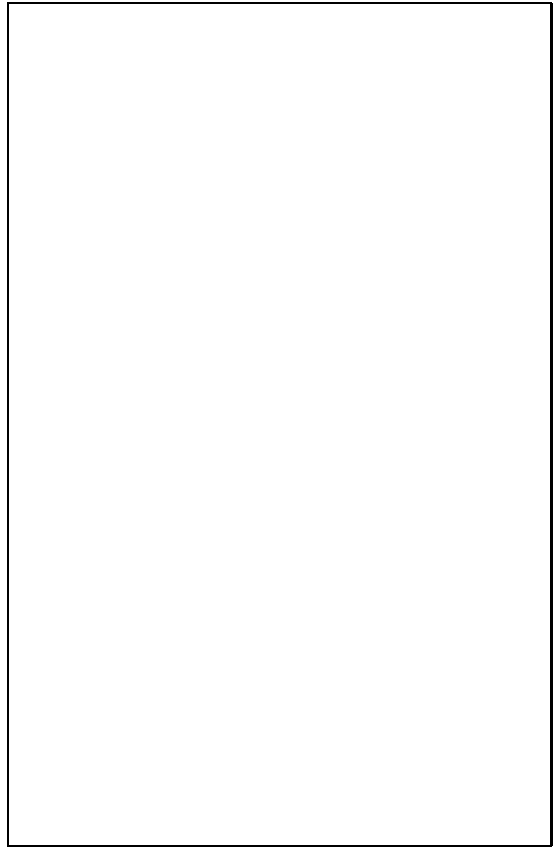

\figureheight{11cm}
\caption{(a)ESR spectra around the anti-level-crossing between O$_2$ and TS$_1$(dotted line). Dashed lines are
eye-guides. (b) Same for O$_1$ and TS$_2$(dotted line). (c) Frequency-field diagram at 1.6 K, where an applied magnetic field is tilted 15 degree from the
$a$-axis in the
$ac$-plane. Dashed lines are
eye-guides.}
\label{fig:7}
\end{figure}

Let us compare the present results with the Raman scattering experiments, where several singlet bound states are found.\cite{pl} 
A clear peak was observed at 30 cm$^{-1}$(900 GHz) in B$_1$ polarization(see Fig. 1 of the Ref.[8]).
In B$_2$ polarization, a peak was also found at 30 cm$^{-1}$ with a shoulder at 24 cm$^{-1}$(720 GHz).
The shoulder appeared also in B$_1$ polarization and a small low energy shift was found at 6 T. Consequently, the shoulder was assigned as one-triplet
excitation being weakly allowed for some defects. 
The energy of TS$_2$ is very close to 30 cm$^{-1}$ and thus TS$_2$ can attribute to the 30 cm$^{-1}$ Raman peak.
Although the energy of TS$_1$ is not far from the 24 cm$^{-1}$ shoulder found in the Raman scattering, these two modes are different each other because
the former is clearly the singlet and the latter was assigned as one-triplet excitation. Consequently the observation of TS$_1$ reveals
that the lowest singlet bound state is located at 646 GHz( 21.5 cm$^{-1}$).
It is notable that the energy is lower than the spin gap of the one-triplet excitation.
This fact indicates that the system is rather close to the quantum
critical point where the energy falls to zero. 

Let us compare the present results with the theoretical investigations on various bound states.\cite{bound,fuk,kn}
As discussed by Totsuka, Miyahara and Ueda, the representations {\it E} and {\it B$_2$} are infrared(also ESR) active and those {\it A$_2$}
and {\it B$_1$} are infrared inactive in {\it D$_{2d}$} symmetry. The one-triplet belongs to
the {\it E} representation.
In Ref. [35], the lowest singlet$_{BS}$ is found to have the {\it A$_2$}$\otimes${\it B$_1$} symmetry(additional small splitting is found
in Ref. [14]).
The second lowest singlet$_{BS}$ coming from the third neighbor pair in Ref. [35] has an infrared active {\it A$_1$}$\otimes${\it B$_2$}
symmetry.
Let us assign tentatively the lowest energy mode and the second lowest energy mode to TS$_1$ and to TS$_2$, respectively.
It is natural to assume that the one-triplet shows the stronger mixing with the infrared active mode.
The strong/weak anti-level-crossings of TS$_2$/TS$_1$ are consistent with the fact that the lower singlet$_{BS}$(TS$_1$) is infrared inactive and that the
upper one(TS$_2$) is infrared active.
However, since only two modes are found in the present experiments, the definite assignment is difficult. A theoretical investigation by considering the
mixing effect is needed for further understanding. Such a calculation may be useful to examine the wave function of the bound states.

\subsection {Triplet bound state}
In Fig. 8, an example of ESR spectrum is shown for {\mib {B}}${\parallel}a$.
Besides one-triplet excitations O$_1$ and O$_2$, eight peaks are found.
These additional peaks are easily classified into triplets$_{BS}$(T$_1$$\sim$T$_5$) and quintets$_{BS}$(Q$_1$$\sim$Q$_3$) by Zeeman shifts in the
frequency-field diagram.
In the inset of Fig. 8, temperature dependence of intensity is plotted for different peaks.
Below 7 K, the intensity increases with decreasing temperature with similar temperature
dependence for all three peaks.
The behavior confirms that the additional peaks are the excitations from the ground state.
\begin{figure}
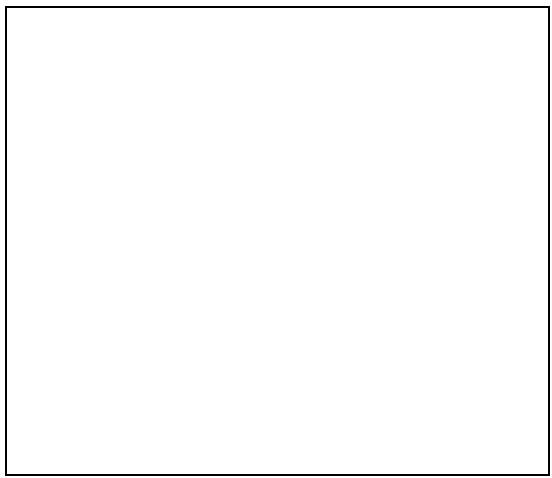

\figureheight{6cm}
\caption{Anisotropy of ESR spectrum at 1024 GHz and at 1.6 K. Dotted
arrows indicate the splitting of O$_1$ and O$_2$. Lines connect corresponding peaks. Temperature dependence of intensity is shown in the inset. See text
for the notation of peaks.}
\label{fig:8}
\end{figure}

A most significant feature in Fig. 8 is the sizable intensity of the bound state excitations.
It is the evidence of the large suppression of the one-triplet process in the present system.
It also supports an idea that a bound state, propagating with correlated motion, has considerable spectral weight.\cite{bound} 
In general, ESR intensity is not a spectral weight itself and is mediated by transition matrix elements mixing the ground and excited
states.
However, the intensity can be compared directly between the one-triplet state and the triplet$_{BS}$.
The argument is based on the fact that a total spin number of the system changes by one in both types of excitations.
From this basic view point, we notice that the magnitude of matrix element may not be so different(at least in the same order) between these two
processes. 
In fact, the intensity ratio of O$_1$+O$_2$ to T$_1$ is comparable with that of the first excited peak to the second one in
neutron scattering(see Fig. 4 of Ref.[16]).\cite{cepas}
This fact supports our presumption that ESR intensity of the triplet states(in both the one-triplet and the bound state) roughly corresponds to the
spectral weight.

Another important finding is that the bound state spectrum splits into many sharp discrete peaks instead of a conventional two-triplet continuum.
It suggests that many different types of bound state is formed for different interactions acting between the distant triplets on dimers.
The fact that these different states do not merge into a single band suggests that the dispersion of the bound states is weak.

Let us evaluate the zero field energy gap of each bound state.
We simply evaluate a gap by extrapolating each mode with a linear equation. 
This procedure may cause non-negligible error, if the level of a bound state has a large bending around zero field as is observed in O$_1$ and O$_2$.
To check this point, we compare a zero field gap obtained for {\mib {B}}${\parallel}a$ with that for {\mib {B}}${\parallel}c$
and found that the difference is at most 10 GHz.
It is much smaller than that of O$_1$ and O$_2$($\sim$40 GHz) and is comparable with experimental errors.
Hence, the above mentioned procedure is justified. The energy gaps of the bound states thus evaluated are listed in Table I. 
We speculate that the anisotropic zero field splittings of the bound states are reduced effectively by the hopping effect or by
the spread of the wave function over distant dimers.
Another finding is that the anisotropy of the intensity of the triplets$_{BS}$ except T$_1$ is similar to that of the one-triplet excitations.
It shows that the origin of the break down of the selection rule may be the same for both cases.
\begin{table}
\caption{Zero field energy gap of the bound states. The notation is identical to that in Figs. 4 (a) and 4(b).}
\label{table:2}
\begin{tabular}{@{\hspace{\tabcolsep}\extracolsep{\fill}}cccc}
\hline	\\
Triplet &Gap (GHz) &Quintet &Gap (GHz)\\
T$_1$	&$1140\pm5$ &Q$_1$	&$1390\pm15$	\\
T$_2$	&$1170\pm5$ &Q$_2$	&$1560\pm15$	\\
T$_3$	&$1190\pm5$ &Q$_3$	&$1600\pm15$	\\
T$_4$	&$1225\pm5$ &Q$_4$	&$1652\pm15$	\\
T$_5$	&$1350\pm5$ &Q$_5$	&$1729\pm15$	\\
T$_6$	&$1220\pm5$ &	\\
T$_7$	&$1264\pm5$ & \\
T$_8$	&$1303\pm5$ &	\\
\hline
\end{tabular}
\end{table}

Several ESR spectra are shown in Figs. 9 (a) and 9 (b) around the one-triplet gap.
The spectra change drastically in a tiny range of frequency with both the shifts and the intensity changes. This behavior may be caused by the
increasing mixing between the triplet$_{BS}$ and the one-triplet state.
Most spectacular feature is found at 736.1 GHz for both {\mib {B}}${\parallel}a$ and {\mib {B}}${\parallel}c$. 
A broad and intense peak, including many different sub-peaks is found in high field side. 
Since the anomaly is found only at the identical frequency irrespective of field direction, it may be caused by the mixing with a horizontal level lying
at this particular energy.
Considering the no-mixing of the triplet$_{BS}$ with TS$_2$(no break is found in Figs. 4 (a) and 4 (b)) and the no-mixing of the one-triplet states
with the new horizontal level, it may not be a singlet bound state.
The $S_z=0$ branch of the one-triplet of zero wave vector point is located at 722 GHz  and it cannot be the candidate of new singlet state.
It is found that, at the (1.5, 0, 0) point, an $S_z=0$ branch of the one-triplet is found
at exactly at this energy by neutron scattering(see Fig. 3 of Ref.[30]). This $S_z=0$ branch may be responsible for the anomaly at
736.1 GHz. Such
mixing between the excitation at zero-wave vector points ( (0, 0, 0) and equivalent points) and
that at $\pi$-wave vector point such as (1.5, 0, 0) is possible by some non-secular terms such as a staggered field. 
Weaker but similar anomalies are found in Fig. 5 as horizontal high intensity lines, which may be caused by the similar origin.
\begin{figure}
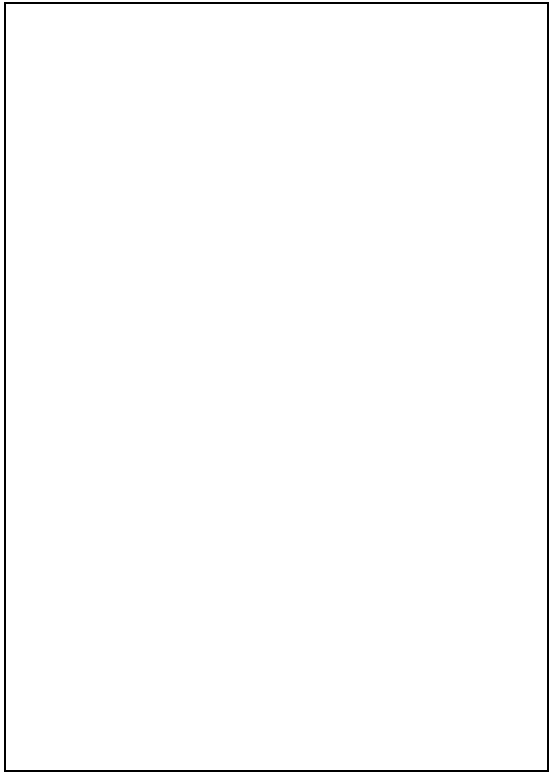

\figureheight{10cm}
\caption{ESR spectra around the one-triplet gap at 1.6 K (a) for {\mib {B}}${\parallel}c$ and (b) for {\mib {B}}${\parallel}a$. Dashed and solid lines are
guides of T$_1$ and T$_4$, respectively. Vertical bars show O$_1$ and O$_2$. B is a peak appearing around
$H_c$.}
\label{fig:9}
\end{figure}

Let us compare the present results with theoretical calculations.
As shown in Table I, all eight bound sates are below the bottom of the two-triplet
continuum (2$E_g$=1444 GHz) and thus these states are stable as a two-triplet bound state. In the reference [\cite{bound}], only five modes(reducing
to three with degeneracy) are found below the bottom. All the mode are infrared active and thus we can attribute at least the five from the all eight
modes to the theoretical ones. In the exact diagonalization result which is more reliable around a quantum critical point, two peaks are found side by
side around 5 meV(1208 GHz). The energy is close to those of T$_1$ and T$_2$. In other theoretical works, the number of states is also less than that found
in the present experiment.\cite{fuk,kn} To have more bound states, the interactions between distant dimers should be taken into account(an inter-layer
coupling may also contribute). Finally we mention the instability of a triplet$_{BS}$ due to the transition to another ground state, proposed in theoretical
work.
\cite{kn} In spite of detailed experiments, no such anomalous behavior is found in the present work. 

\subsection {Quintet bound state}
In the following, we show the first observation of a quintet$_{BS}$ in the system.
The nature of the magnetization appearing around the $H_c$ is also discussed.
As shown in Fig. 8, the intensity of a quintet$_{BS}$ is weaker roughly one order of magnitude, which is due to the higher order excitation
process of the total spin change:$\Delta${\it S}=2. 
Five quintet$_{BS}$ modes are found in the present work, which are easily identified for the typical double Zeeman shift.
A zero energy gap of a quintet$_{BS}$ has been extrapolated from the experimental data.
Since the number of data points is limited, an error of the gap is about 15 GHz as listed in Table I.
The anisotropy of the gap is not found within the experimental error.

It is notable that only Q$_1$ is below the two-triplet continuum threshold starting at 2$E_g$=1444 GHz, while the other four modes are located above the
threshold. Since the energy of Q$_1$ is below the threshold, it is assigned as a quintet$_{BS}$ made up of two-triplet particles(four spins).
Let us discuss the nature of other high energy quintets$_{BS}$.
In the isolated dimer limit of four spins, the triplet$_{BS}$ and quintet$_{BS}$ are separated by the effective antiferromagnetic interaction
acting among the four spins. In this limit, the higher energy of Q$_2\sim$Q$_5$ can attribute to this interaction and those peaks are considered as
two-triplet bound state.
This interpretation has a difficulty because a multi-particle bound state is unstable when the energy is above the corresponding incoherent continuum in
general. It is totally unclear if the triplet and the quintet sectors of the bound state can separate clearly from the continuum spreading above 2$E_g$.
A more natural idea is to assign Q$_2\sim$Q$_5$ to three-triplet bound state and in fact such three triplet state is found in theoretical calculations,
which is shown in later.
The existence of three-triplet bound state shows the significant contribution of higher order
excitations.

The existence of Q$_1$ below the threshold causes the peculiar feature that the quintet$_{BS}$ touches the ground state before the closing of
one-triplet  gap. In spite that a finite mixing between the ground and excited states modifies this simple situation to the more complex one, a
characteristic contribution of a quintet is found experimentally.
The enlarged spectra of Figs. 4 (a) and 4 (b) around the $H_c$ are shown in Figs. 10 (a) and 10 (b).
It is noticed that ESR modes bend around the $H_c$ and the behavior is much different between {\mib {B}}${\parallel}c$ and {\mib {B}}${\parallel}a$. 
In fact, the antistory of Q$_1$ cannot be normalized by the $g$-value and thus is originated by the anisotropic mixing term.
This anisotropy of ESR spectra should correspond to the anisotropy of the magnetization curves.
It is because the magnetization around the $H_c$ is caused by the mixing between the ground and the excited states.
In another word, the magnetization should be dominated by the wave function component
of the modes enhanced by the mixing.
Keeping this point in mind, let us examine the magnetization and the frequency-field diagram carefully.

For {\mib {B}}${\parallel}c$, the magnetization becomes finite above 15 T and at the same time O$_2$ starts to deviate from a straight line.
This fact shows that the one-triplet component dominates the magnetization.
The intensity of Q$_1$ is enhanced around 20 T presumably for the weak anti-level-crossing with O$_1$.
Around 21 T, Q$_1$ disappears and this fact indicates that the quintet component in the magnetization is small.
After the anti-level-crossing, O$_1$ bends considerably and loses the intensity gradually.
It is noticed that O$_2$ shows a turn around like behavior above 20 T, which indicates the strong mixing with the ground state.
On the other hand O$_1$ goes down again above 26 T and disappears suddenly at the jump of magnetization to the \(\frac18\)-plateau as shown in Fig. 4
(a). The origin of the flat part of O$_1$ around 25 T might be a anti-level-crossing with a new singlet$_{BS}$, however, a further investigation is
necessary for this point. The strong mixing of O$_2$ with the ground state indicates the dominant contribution of the one-triplet component to the
magnetization. It is also noticed that Q$_3$ is enhanced with a large rounding above 20 T, which shows the mixing of three-triplet state with the ground
state. The contribution of different components to the magnetization indicates that the magnetization is not uniform and
homogeneous. It is much different from a usual spin gap system in which a magnetization consists of homogeneously distributed triplets. 
This characteristic behavior may be caused by the localization of excited states in the present system.
Recently, a homogeneous internal field is found by NMR below the \(\frac18\)-plateau, however the difference is presumably due to the
low observation frequency in NMR, in which the distribution of an internal field is averaged.\cite{nmr}
\begin{figure}
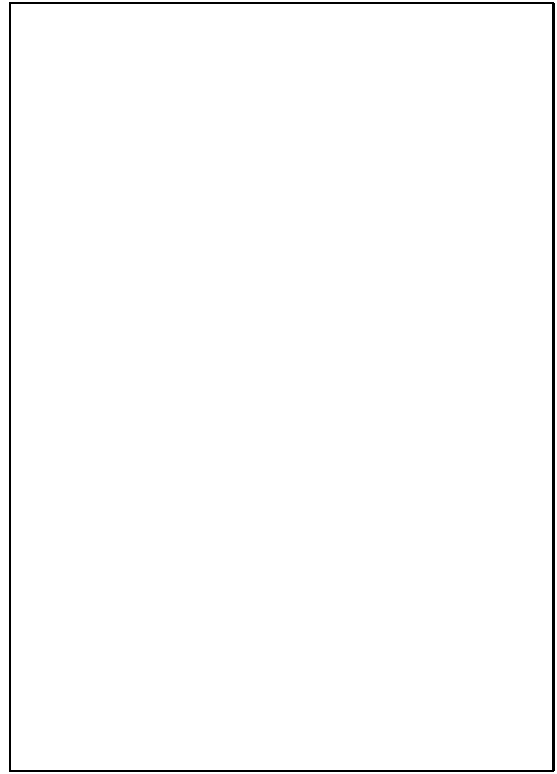

\figureheight{10cm}
\caption{The low energy part of ESR spectra around $H_c$ (a) for {\mib {B}}${\parallel}c$ and (b) for {\mib {B}}${\parallel}a$. The dashed lines are
eye-guides. The vertical bar denotes Q$_1$.}
\label{fig:10}
\end{figure}

 Let us discuss the case of {\mib {B}}${\parallel}a$ shown in Fig. 10 (b).
While the behavior of O$_1$ is similar to that in {\mib {B}}${\parallel}c$, both Q$_1$ and O$_2$ continuously go down toward the ground state.
The enhancement of Q$_1$ and the disappearance of O$_2$ indicates that the quintet$_{BS}$ component dominates the magnetization.
Other higher energy quintet$_{BS}$ states also exhibit enhancements and bendings, which indicate the strong mixing of the quintets$_{BS}$ with the ground
state. An observation of a small break of O$_2$ in {\mib {B}}${\parallel}a$(see Fig. 4(b)) and at 22 T may be caused by the anti-level-crossing with unknown
singlet$_{BS}$ at 100 GHz, which is also found in {\mib {B}}${\parallel}c$.
  
As shown above, the magnetization below the \(\frac18\)-plateau contains different types of states and does not consist of homogeneously
distributed triplets. A dominant contribution of a quintet$_{BS}$ is found for {\mib {B}}${\parallel}a$.
It is an interesting point, if a quintet$_{BS}$ coupled by an effective repulsive interaction shows a condensation at $H_c$, which was proposed by Momoi
and Totuska.\cite{mt}
Experimentally the domination of the quintet$_{BS}$ is found clearly in the present work. However, it may be a quenched and localized state rather than a
condensation. In the case of a condensation, a magnetization should not contains a several different states as observed in the present case.
The localization of excited states and the resulting heavy effective mass of a quintet$_{BS}$ may prevent the condensation of a quintet$_{BS}$.

Finally we compare the present results with the theoretical calculation of the quintet$_{BS}$.
In the reference [\cite{bound}], the lowest mode has the {\it A$_2$}$\otimes${\it B$_1$} symmetry, which is infrared inactive.
Since a quintet$_{BS}$ shows very weak intensity, an infrared inactive mode should not be excluded from the candidate.
The fact that the number of two-triplet quintet$_{BS}$ is only a few is consistent with the present result.
However, a large  number of a three-triplet quintet$_{BS}$ found in the experiment calls for further theoretical investigations.

\section {High Field Phase}
In this section, we discuss ESR signals observed above $H_c$ and their concerns with the magnetization plateaux.
In this region of magnetic field, the most of the intensity concentrates into the signals B and M as shown in Figs. 3 (a) and 3 (b). 
These two signals alternate at the magnetization jump to the \(\frac18\)-plateau as shown in Fig. 11.
The position of M is identical with that of the corresponding paramagnetic resonance and the extrapolation of M crosses the origin of the frequency-field
diagram for both {\mib {B}}${\parallel}c$ and {\mib {B}}${\parallel} a$ as shown in Figs. 4 (a) and 4(b).
On the other hand, B shows the asymptotic behavior to the paramagnetic line.
When a spin gap is closed, a 3D ordered-state or a disordered ground state dominated by gap-less triplets appears.
In the former, a field variation of a spin wave is observed as an antiferromagnetic resonance(AFMR).
In the latter, the Zeeman splitting of a triplet causes ESR at an identical resonance field with that of a paramagnetic resonance. 
In the disordered state, the strong correlation among the spins does not cause a shift of resonance field.
It is because the spin operator commutes with the isotropic exchange coupling. \cite{OA}

The behavior of M shows that the signal is the ESR of a disordered state.
It is noticed that M continues at \(\frac14\)-plateau and a low-energy excitation is absent.
Such low energy discrete excitation is observed at \(\frac13\)-plateau, which corresponds to the creation of a new triplet into a super-lattice structure at
a plateau.\cite{kimura} These features can be understood as the analogy of Ising spin cluster excitation.
When Ising anisotropy is added gradually to a Heisenberg spin system, a spin wave excitation shows a crossover to a discrete spin cluster excitation.
In the present system, Ising anisotropy corresponds to the hardness of a plateau.
If the plateau is soft, a creation of single triplet(spin cluster like excitation) does not occur and a perturbation by the radiation transmits through
the transverse components of spins as a wave. This excitation is similar to a spin wave excitation because it is likely a gap-less Goldstone mode.
Momoi and Totuska proposed that a "supersolid" state appears between the plateaux, in which a spin density wave and superfluid of spins coexist.\cite{mt}
The latter component gives rise to a gap-less ESR. The present result indicates that a superfluid component(gap-less mode) is significant
at \(\frac14\)-plateau and that the plateau is not completely static at 1.6 K. Recently, the static super lattice structure is found by NMR in the
\(\frac18\)-plateau at 30 mK.\cite{nmr} For \(\frac14\)-plateau, a static super structure may also appear if we measure in much lower temperature below 1.6
K. This point remains for further experimental investigations.

Although the behavior of M is rather conventional, B shows unusual features:(1)location above the paramagnetic line, (2)non-linear field dependence,
and (3)disappearance at the jump to the \(\frac18\)-plateau.
Considering these features, we propose two possible origins of B:(a)AFMR, (b)transition ${S_z=1 \to S_z=0}$ in the lowest triplets.
First we discuss the possibility of AFMR.
The asymptotic behavior of B matches with that of AFMR mode when a magnetic field is perpendicular to the plane of spin laying.
In AFMR, a deviation from the paramagnetic line is caused by anisotropy and thus the deviation should change for different field orientation.
However, the deviation of B is nearly equal between {\mib {B}}${\parallel}c$ and {\mib {B}}${\parallel}a$.
The most fundamental difficulty of the interpretation by AFMR is that no evidence of 3D-ordering is found below the \(\frac18\)-plateau.
Actually, in NMR measurement, no static internal field is found in this field range, which shows non-existence of 3D-order.\cite{nmr}
A possible interpretation is that a quasi static order takes place and it fluctuates with the frequency between ESR and NMR.
In another word, the system is considered to be dynamical in the time window of NMR and to be static in that of ESR.
However, it is not clear if such situation is realized. 

Next we discuss the possibility of the case (b). In the most simple picture of a isolated dimer system, the $S_z=1$ level is below the singlet ground
state above $H_c$ as shown in the inset of Fig. 11. 
The transition ${S_z=1 \to S_z=0}$ shown by the arrow($\nu_2$) represents ESR of a disordered magnetic state and the resonance field is identical with that
of a paramagnetic resonance.
In the present system, the level scheme is modified for the anti-level-crossing as shown by the dotted lines.
The transition ${S_z=1 \to S_z=0}$ also changes to that of the dashed arrow($\nu_1$).
The break by the anti-level-crossing causes the deviation of the transition from that of a simple triplet.
The signal B may be assigned as the transition $\nu_1$. 
Qualitatively, the features (1) and (2) mentioned above are consistent with this interpretation. 
The strong intensity of B is also consistent with the picture because it is the allowed transition from the ground state.
However, a quantitative agreement is difficult with the simple anti-level-crossing model of the inset. 
In the simple model shown in the inset, the relation ${\nu_1}$=${\nu_2}$+${\nu_3}$ should hold at each field, where ${\nu_3}$ is the
transition ${S_{GS}=0 \to S_z=1}$(note that arrows in the inset are shifted to prevent the overlapping). It is because of the anti-symmetry of the
anti-level-crossing around the $H_c$. However, this relation is not found among O$_1$, O$_2$ and B as can be seen clearly in Figs. 4 (a) and 4 (b). Namely
the deviation of B is rather independent of the field direction, O$_1$ and O$_2$ show the anisotropic behaviors.
As shown above, both proposal are not sufficient to explain the behavior of the signal B.
The nature of the state below the \(\frac18\)-plateau seems to be unconventional, which requires further microscopic investigations by
using different types of probes.

\begin{figure}
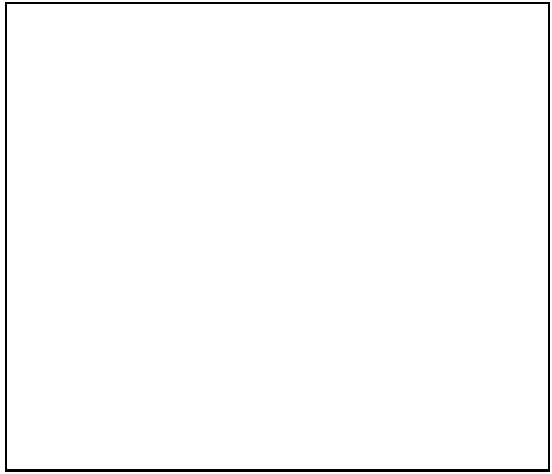

\figureheight{6cm}
\caption{The ESR spectra around the \(\frac18\)-plateau for {\mib {B}}${\parallel}c$ at 1.6 K. The thick and thin dashed lines indicate the paramagnetic
line and the field of the magnetization jump to the \(\frac18\)-plateau, respectively. The dotted line is an eye-guide for the signal B. The inset
shows the schematic energy level of an isolated $S=1/2$ antiferromagnetic dimer. ${S_{GS}=0}$ is the singlet ground state. For others, see text.}
\label{fig:11}
\end{figure}

\section {Summary}
To be summarized, the following important features are found by the detailed investigation of the magnetic excitation spectrum up to 1.3 THz and up to 40
T.
(1)Difference of the symmetry between the nearly degenerated one-triplet excitations O$_1$ and O$_2$ are found in the field dependences and in the
anti-level-crossing with the singlet$_{BS}$  and the ground state.

(2)Two-singlet bound states are found by the anti-level-crossing with the one-triplet excitation. The one of those is located below the spin gap of the
one-triplet, which indicates the proximity to the quantum critical point of the present system.

(3)Several triplet bound states are found and the zero field energy gaps are all below the two-particle continuum threshold.

(4)The gap of the lowest quintet$_{BS}$ is found to be below the two-particle continuum threshold and it crosses with the one-triplet excitation
before the closing of the normal one-triplet spin gap. Some of the quintet$_{BS}$ modes are assigned as a three-triplet bound state. 

(5)The finite magnetization appearing below the \(\frac18\)-plateau contains various types of magnetic components caused by the anisotropic mixing and the content changes
with the field direction.

(6)Two models are proposed for the unusual ESR mode B observed below the magnetization jump to the \(\frac18\)-plateau.

(7)The domination of gap-less ESR signal M indicates that the \(\frac14\)-plateau is rather soft compared to the \(\frac13\)-plateau.

(8)Qualitative interpretation is made for the selection rules in the one-triplet excitation by considering intra-dimer DM-interaction, inter-dimer
DM-interaction and a staggered field. 

The present work exhibits various novel features of higher order multiple-particle excitations in 	a quantum spin gap system.

\section*{Acknowledgements}

The authors would like to thank T. Sakai, T, Ziman, O. C$\acute e$pas, K. Totsuka and Y. Fukumoto for valuable discussions.
This work was partly supported by a Grant-in-Aid for Scientific Research on Priority Areas from the Ministry of Education, Culture, Sports, Science and
Technology.

\end{document}